\documentclass[12pt,a4paper,final]{iopart}

\usepackage{iopams}

\expandafter\let\csname equation*\endcsname\relax

\expandafter\let\csname endequation*\endcsname\relax

\usepackage[utf8]{inputenc}  %Input what you want e.g., é, ł, a, ü
\usepackage[T1]{fontenc}     %Output what you want e.g., é, ł, a, ü
\usepackage[british]{babel}  %Do hyphenation according to british english
\usepackage[sc,osf]{mathpazo}
\usepackage[scaled=0.86]{berasans}  % URL font that go well wtih palatino
\usepackage[scaled=1.03]{inconsolata} %Monospace font
\usepackage[colorlinks,citecolor=blue,linkcolor=magenta,urlcolor=blue]{hyperref}  %Hyperlinks (pink, green, blue)
\usepackage{graphicx} % Package to insert exteral figures
\usepackage{iopams}
\usepackage{dsfont}
\usepackage{xspace}  %Useful to add space in macros
\usepackage{centernot}
\usepackage{pgfplots}  %Useful to plot graphs with latex
\usepackage{lmodern}  %Joe's font
\usepackage[sc,osf]{mathpazo}
\usepackage[scaled=0.86]{berasans}  % URL font that goes well wtih palatino
\usepackage[scaled=1.03]{inconsolata} %Monospace font
\usepackage[colorlinks,citecolor=blue,linkcolor=magenta,urlcolor=blue]{hyperref}  %Hyperlinks (pink, green, blue)
\usepackage[babel]{microtype}  %Improves text justification
\usepackage{amsmath,amssymb,amsthm,bm,mathtools,amsfonts,mathrsfs,bbm} %Usefull math packages
\newcommand{\bra}[1]{\left\langle #1 \right|}
\newcommand{\ket}[1]{\left| #1 \right\rangle}

\newcommand{\ba}{\begin{eqnarray}}
\newcommand{\ea}{\end{eqnarray}}
\begin{document}
\title{Entanglement without hidden nonlocality}

\author{Flavien Hirsch}
D\'epartement de Physique Th\'eorique, Universit\'e de Gen\`eve, 1211 Gen\`eve, Switzerland

\author{Marco T\'ulio Quintino}
D\'epartement de Physique Th\'eorique, Universit\'e de Gen\`eve, 1211 Gen\`eve, Switzerland

\author{Joseph Bowles}
D\'epartement de Physique Th\'eorique, Universit\'e de Gen\`eve, 1211 Gen\`eve, Switzerland

\author{Tam\'as V\'ertesi}
Institute for Nuclear Research, Hungarian Academy of Sciences, H-4001 Debrecen, P.O. Box 51, Hungary

\author{Nicolas Brunner}
D\'epartement de Physique Th\'eorique, Universit\'e de Gen\`eve, 1211 Gen\`eve, Switzerland

\date{\today}  %Date today

\begin{abstract}
We consider Bell tests in which the distant observers can perform local filtering before testing a Bell inequality. Notably, in this setup, certain entangled states admitting a local hidden variable model in the standard Bell scenario can nevertheless violate a Bell inequality after filtering, displaying so-called hidden nonlocality. Here we ask whether all entangled states can violate a Bell inequality after well-chosen local filtering. We answer this question in the negative by showing that there exist entangled states without hidden nonlocality. Specifically, we prove that some two-qubit Werner states still admit a local hidden variable model after any possible local filtering on a single copy of the state. 
\end{abstract}

\maketitle

Nonlocality is one of the most startling predictions of quantum mechanics. It allows two distant observers to obtain experimental statistics that cannot be described by any classical common cause (given a few reasonable physical assumptions) \cite{bell64,brunner_review}. Recently confirmed in loophole-free experiments \cite{hensen15,giustina15,shalm15} nonlocality has been proven to be useful for many tasks, such as device-independent cryptography \cite{acin07} and randomness certification \cite{pironio10,colbeck_thesis}. 

This effect is enabled by the genuinely quantum phenomenon of entanglement. However, it is still unclear which entangled quantum states lead to nonlocality \cite{augusiak_review}. While for pure states it is known that all entangled states can display nonlocal correlations \cite{gisin91,popescu92,guhne16}, mixed states exhibit a more intricate behaviour, as initially shown by Werner \cite{werner89}. Namely, there exist entangled mixed states that never lead to nonlocality when submitted to any local measurements, even taking general POVMs into account \cite{barrett02}. Following earlier results of \cite{toth05,augusiak14}, this phenomenon has recently been shown to hold true in the general multipartite case as well: for any number of parties, there exist genuinely multipartite entangled states which admit a fully local hidden-variable (LHV) model \cite{bowles16}. 

However, these results have been derived in the scenario considered initially by Bell, i.e., in each run of the experiment, non-sequential local measurements are performed on a single copy of an entangled state. Going beyond this standard Bell scenario allows one to reveal the nonlocality of some entangled states which admit a LHV model (in the standard Bell scenario). For instance, one could allow for local filtering, i.e. local filters applied by each party \textit{before} the standard Bell test, hence being a pre-processing of the entangled state. This was first proposed by Popescu \cite{popescu95}, who concluded that some entangled Werner states which admit a LHV model (for all projective measurements) display some `hidden nonlocality', that is, violate a Bell inequality after well-chosen local filters. This phenomenon has been shown to exist even for entangled states admitting a LHV model for general measurements (POVMs)  \cite{hirsch13}. Hence, local filtering allows one to reveal the nonlocality of some entangled states which are always local in the standard Bell scenario.

Following Ref. \cite{popescu95}, several aspects of local filtering in Bell tests have been discussed. For the two-qubit case, Ref. \cite{verstraete01} studied how local filtering can increase entanglement and it was shown that local filtering can `activate' CHSH-violation \cite{gisin96,verstraete02}, for which a necessary and sufficient condition was derived \cite{pal14}. Ref. \cite{masanes05} generalized hidden nonlocality to many copies and showed a strong link to entanglement distillability. Ref. \cite{masanes08} showed that all entangled states display some kind of hidden nonlocality, in the sense that any entangled state can help to activate the CHSH violation of another entangled state. Refs \cite{zukowski98,gallego14} discussed the general scenario of Bell tests with sequential measurements, of which local filtering is a particular case. Finally, local filtering was shown to reveal genuine multipartite nonlocality \cite{bowles16}. Altogether local filtering has been shown to be a powerful way of activating nonlocality from entangled states which admit LHV models. 

A natural question is therefore whether local filtering can reveal the nonlocality of \textit{all} entangled states. That is, do all entangled states display hidden nonlocality? Here we answer this question in the negative, by showing that some entangled states cannot exhibit hidden nonlocality, considering the scenario of Popescu. Specifically, we show that some entangled two-qubit Werner states admit a LHV model after local filtering on a single copy of the state. Our model takes into account any local filters, and holds for all local POVMs performed on the state after filtering. Our result can also be interpreted as follows: some entangled two-qubit Werner states cannot violate any Bell inequality, even after arbitrary stochastic local operations and classical communication (SLOCC) performed before the Bell test on a single copy of the state\cite{masanes05}. We conclude with some open questions.

\section{Preliminaries}

\subsection{Bell nonlocality and quantum steering}

Consider two distant observers, say Alice and Bob, sharing an entangled quantum state $\rho$. Alice performs one measurement chosen in a set $\{M_{a|x} \}$ ($M_{a|x}\geq 0$ and $\sum_a M_{a|x} = \mathds{1}$), and Bob performs a measurement chosen in a set $\{M_{b|y}\}$ (with similar conditions). Given that they choose the measurements labelled by $x$ and $y$, respectively, the resulting statistics are given by
\begin{align} \label{pQ}
p(ab|xy) = \Tr ( M_{a|x} \otimes M_{b|y} \; \rho ).
\end{align}
The state $\rho$ is said to be local for measurements $\{M_{a|x} \}$, $\{M_{b|y}\}$ if the distribution \eqref{pQ} admits a Bell-local decomposition
\begin{align}
\label{LHV} p(ab|xy) &= \int \pi(\lambda) \; p_A (a|x,\lambda ) \; p_B (b|y,\lambda) \; d\lambda .
\end{align}
That is, the quantum statistics can be reproduced using a LHV model consisting of a shared local variable $\lambda$, distributed with density $\pi(\lambda)$, and local response functions given by distributions $p_A (a|x,\lambda )$ and $p_B (b|y,\lambda)$. If for some sets of measurements $\{M_{a|x} \}$ and $\{M_{b|y}\}$ a decomposition of the form \eqref{LHV} cannot be found, the distribution $p(ab|xy)$ violates (at least) one Bell inequality \cite{brunner_review}. In this case, we conclude that $\rho$ is nonlocal for the measurements $\{M_{a|x} \}$ and $\{M_{b|y}\}$.

Another concept that will be useful here is that of EPR-steering \cite{wiseman07}; see \cite{paul_review} for a review. It is a weaker form of nonlocality which captures the fact that if Alice makes a measurement on her half of the state $\rho$ she remotely steers Bob's state. This nonlocal effect can be detected in the statistics of the experiment if Bob measures his part of the state as well. Specifically, if Alice and Bob perform measurements $\{M_{a|x} \}$ and $\{M_{b|y}\}$, respectively, we say that $\rho$ is `unsteerable' (from Alice to Bob) if
\begin{align}
\label{LHS} p(ab|xy) &= \int \pi(\lambda) \; p_A (a|x,\lambda ) \; \Tr ( M_{b|y} \sigma_\lambda) \; d\lambda .
\end{align}
That is, the quantum statistics can be reproduced by a so-called local hidden state model (LHS), where $\sigma_\lambda$ denotes the local quantum state and $p_A (a|x,\lambda )$ is Alice's response function. If such a decomposition cannot be found, $\rho$ is said to be `steerable' for the set $\{M_{a|x}\}$; note that one would usually consider here a set of measurements $M_{b|y}$ that is tomographically complete, and thus focus the analysis on the set of conditional states of Bob's system
\begin{align}
\sigma_{a|x} = \Tr_A (M_{a|x} \otimes \mathds{1} \; \rho),
\end{align}
referred to as an assemblage. Note that any LHS model is also a LHV model, although the converse does not hold. 

If a state $\rho$ admits a decomposition \eqref{LHV} for all measurements $\{M_{a|x} \}$ and $\{M_{b|y}\}$ we say that $\rho$ is local, i.e. $\rho$ admits a LHV model. Similarly if $\rho$ admits a decomposition \eqref{LHS} for all measurements $\{M_{a|x} \}$ we say that $\rho$ is unsteerable. With these definitions we have that entanglement, steering and nonlocality are strictly different concepts, even taking all possible POVMs into account. More precisely, one can show that there exist entangled states which are unsteerable (hence local) states, as well as entangled local states which are steerable. Indeed, Werner showed that some entangled quantum states admit a LHS model \eqref{LHS} for \textit{all} projective measurements \cite{werner89}. This result was later extended to general POVMs \cite{barrett02}. Similarly, certain steerable states were shown to admit a LHV model for projective measurements \cite{wiseman07}, and the same hold considering general POVMs \cite{quintino15}. LHV and LHS models for different classes of entangled states were also constructed, see e.g. \cite{acin06,almeida07,hirsch13,sania14,bowles14,bowles15,bowles15b,nguyen16}.

\begin{figure}[b!] \begin{center}
\includegraphics[width=0.85\columnwidth]{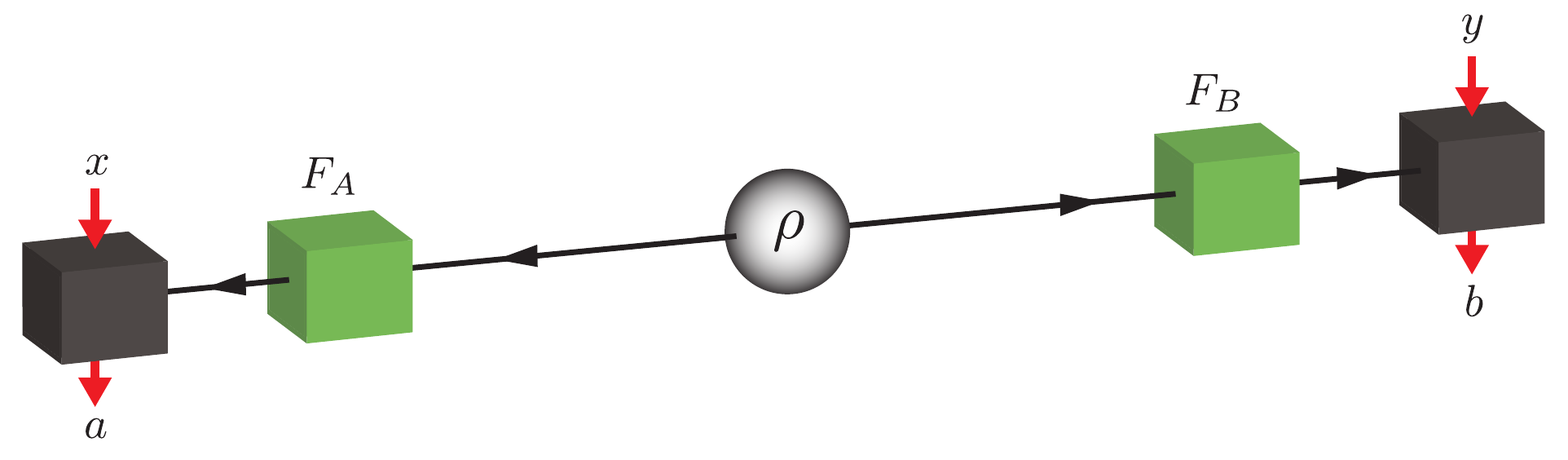}
\caption{The hidden nonlocality scenario: Alice and Bob share an entangled state $\rho$ and perform local filters $F_A$ and $F_B$, respectively. When the filtering is successful, they perform a standard Bell test. If the resulting statistics $p(ab|xy)$ violate a Bell inequality, the state $\rho$ displays hidden nonlocality. Here we ask if this effect is possible for all entangled states, and show that this is not the case.}
\end{center}
\end{figure}

\subsection{Hidden nonlocality}

One could conclude from the above that local entangled states are somehow classical, as they can always be replaced by classical variables $\lambda$ (with no noticeable difference in any Bell experiment). Nevertheless the nonlocality of some local entangled quantum states can in fact be revealed in more complex ways than the traditional Bell scenario. As first imagined by Popescu \cite{popescu95}, one could submit a quantum state to a \textit{sequence} of measurements. Indeed in quantum mechanics a measurement generally changes the state, leading to different statistics when a second round of measurements is applied. Note that a measurement does not necessarily break the entanglement of the quantum state. On the contrary, for a given measurement outcome, entanglement can be increased.

The simplest way to implement this idea is that of local filtering: Alice and Bob first perform local filters on their shared state $\rho$, given by a set of Kraus operators $\mathbb{F}_A= \{ F_A,\bar{F_A} \}$ and  $\mathbb{F}_B= \{ F_B, \bar{F_B} \}$, where $F_{A}^{\dagger} F_A + \bar{F_A}^{\dagger} \bar{F_A} = \mathds{1}$ and similarly for Bob. Alice and Bob keep the post-filter state only when $\rho$ `passes' the filter, meaning Alice obtained the outcome corresponding to $F_A$ and Bob the outcome corresponding to $F_B$. The state they hold in that case is given by
\begin{align}
\label{filtering} \rho'= \frac{F_A \otimes F_B \; \rho \; F_A^{\dagger} \otimes F_B^{\dagger}}{\tr(F_A \otimes F_B \; \rho \; F_A^{\dagger} \otimes F_B^{\dagger})} .
\end{align}
In terms of state transformation the operation which transforms $\rho$ into $\rho'$ can be seen as a stochastic local operation (SLO). Note that $F_A$ and $F_B$ are any linear operators acting on $\mathcal{H}_A$ (respectively $\mathcal{H}_B$) and can in particular increase or decrease the dimension of the Hilbert space of the quantum state.

Next Alice and Bob can perform a standard Bell test on $\rho'$, that is they perform a second round of local measurements $M_A=\{M_{a|x}\}$ and $M_B=\{M_{b|y}\}$. Repeating the process many times one can access the statistics $p(ab|xy) = \Tr ( M_{a|x} \otimes M_{b|y} \; \rho' )$, and check whether it admits a Bell-local decomposition \eqref{LHV}. This scenario is illustrated in Figure 1.

Can $\rho'$ be nonlocal although $\rho$ admits a LHV model (in the standard Bell scenario)? Ref \cite{popescu95} showed that indeed this can be the case for certain Werner states admitting LHV models for projective measurements, while Ref. \cite{hirsch13} extended this result by considering a state $\rho$ admitting a LHV model for general POVMs. Hence there are entangled states that cannot lead to nonlocality in the standard Bell-scenario (even taking general POVMs into account), but nevertheless violate a Bell inequality after local filtering. 

These examples open the question of whether all entangled states can lead to hidden nonlocality. That is, for any entangled state $\rho$, can we find local filters $\mathbb{F}_A$ and $\mathbb{F}_B$ such that the resulting state $\rho'$ is nonlocal? We answer this question in the negative by constructing an explicit counter-example.

%A VOIR... We also remark that $\rho$ violates a particular Bell inequality after local filtering if and only if $\rho$ also violates the same Bell inequality after deterministic Local Operations and Classical Communications (LOCC) \cite{masanes05}, providing a different interpretation of hidden nonlocality and our result. 

\section{Main result}

Consider the two-qubit Werner state:
\begin{align} \label{werner}
\rho_W(\alpha) = \alpha \ket{ \phi^{+}}\bra{ \phi^{+}}+(1-\alpha)\mathds{1}/4
\end{align}
where $\ket{ \phi^{+}}=(\ket{00}+\ket{11})/\sqrt{2}$ is the maximally entangled two-qubit state and $\mathds{1}/4$ is the maximally mixed state. The state $\rho_W(\alpha)$ is entangled if and only if $\alpha > 1/3$. While Werner originally constructed a LHS model for $\alpha = 1/2$ and projective measurements, this was later extended. Indeed, local models were presented, for all projective measurements and $\alpha \lesssim 0.66$ \cite{acin06}, and for POVMs for $\alpha \leq 5/12$ \cite{barrett02}. The state is steerable for $\alpha > 1/2$ \cite{wiseman07}, and nonlocal for $\alpha > 0.7055$ \cite{vertesi08,hua15}.

Our main result is that $\rho_W(\alpha)$ remains local, considering arbitrary POVMs, after \textit{any} local filtering for $\alpha \lesssim \alpha_c = 0.3656$. Hence the entangled state $\rho_W(\alpha_c)$ displays no hidden nonlocality. This can be formalized with the following theorem:

{\bf Theorem 1.}
 For $\alpha \leq \alpha_c$ the state
\begin{align}
 \rho'= \frac{F_A \otimes F_B \; \rho_W(\alpha) \; F_A^{\dagger} \otimes F_B^{\dagger}}{\Tr(F_A \otimes F_B \; \rho_W(\alpha) \; F_A^{\dagger} \otimes F_B^{\dagger})}
\end{align}
is local for all POVMs. Here, $F_A$ and $F_B$ represent any possible local filters; $F_A , F_B : \mathcal{H}^2 \rightarrow \mathcal{H}^d$.

\textit{Proof.} We will proceed in two steps. First we characterize the filtered state when only Alice applies a local filter. Then we show that this state remains local over all operations applied locally by Bob.

Consider again the Werner state $\rho_W(\alpha)$ defined in \eqref{werner}. Alice applies a local filtering $\mathbb{F}_A= \{ F_A,\bar{F_A} \}$. If $\rho_W(\alpha)$ passes the filter, Alice and Bob hold the state $\rho_{F_A}$ given by:
\begin{align}
 \rho_{F_A} = \frac{F_A \otimes \mathds{1} \; \rho_W(\alpha) \; F_A^{\dagger} \otimes \mathds{1}}{\Tr(F_A \otimes \mathds{1} \; \rho_W(\alpha) \; F_A^{\dagger} \otimes \mathds{1}) }
\end{align}
where $\mathds{1}$ is the identity operator in Bob's Hilbert space. One can show that this state is unsteerable from Alice to Bob (for all POVMs) if and only if the state
\begin{align} \label{Mafalda}
\rho(\alpha,\theta) = \alpha \ket{ \psi_{\theta}} \bra{ \psi_{\theta}} + (1-\alpha) \rho_A \otimes \mathds{1}/2
\end{align}
is unsteerable from Alice to Bob (POVMs), for all $\theta \in [0,\pi/4]$. Here, $\ket{ \psi_{\theta}} = \cos\theta \ket{00} + \sin\theta \ket{11}$ is the partially entangled state  and $\rho_A = \Tr_B(\ket{ \psi_{\theta}} \bra{ \psi_{\theta}})$ its partial trace. 

To prove this claim consider first the unnormalized filtered state
\begin{align}
 \rho_{F_A} = F_A \otimes \mathds{1} \; \rho_W(\alpha) \; F_A^{\dagger} \otimes \mathds{1}	.
\end{align}
 Using the singular value decomposition one can write $F_A = U D V^{\dagger}$, where $U, V$ are unitary matrices and $D$ is diagonal and positive. Note that since $F_A$ is a $d \times 2$ matrix, $U$ is $d \times d$, $D$ is $d \times 2$ and $V$ is $2 \times 2$. We thus have
\begin{align}
 \rho_{F_A}   = U D V^{\dagger} \otimes \mathds{1} \; \rho_W(\alpha) \; V D^T U^{\dagger} \otimes \mathds{1}  = U \otimes \mathds{1} (D V^{\dagger} \otimes \mathds{1} \rho_W(\alpha) V D^T \otimes \mathds{1}) U^{\dagger} \otimes \mathds{1} .
\end{align}
We can then use the fact that if the state $\rho$ is unsteerable so is $U_A \rho U_A^{\dagger}$ (for an arbitrary unitary $U_A$ acting on Alice's subspace) as the statistics coming from a measurement $\{M_a\}$ is $\tr(M_a U_A \rho U_A^{\dagger} )= \tr(U_A^{\dagger} M_a U_A \rho) = \tr(M_a' \rho)$, where $\{M_a'\}$ is another (valid) measurement. We can therefore focus on the unormalized state
\begin{align}
 D V^{\dagger} \otimes \mathds{1} \; \rho_W(\alpha) \; V D^T \otimes \mathds{1}   .
\end{align}
By the same observation as above we can apply any unitary $U_B$ on Bob's side. Choosing $U_B=V^T$ we get
\begin{align}
  D \otimes \mathds{1} ( V^{\dagger} \otimes V^T \; \rho_W(\alpha) \; V \otimes V^*) D^T \otimes \mathds{1}
 = D \otimes \mathds{1} \rho_W(\alpha) D^T \otimes \mathds{1}
\end{align}
using the $U \otimes U^{*}$ symmetry of $\rho_W(\alpha)$. Finally, note that the normalization $\Tr(\rho_{F_A}) = \Tr(D^T D)/2$ is independent of $\alpha$. The one-side filtered Werner state $\rho_{F_A}/\Tr(\rho_{F_A})$ is thus equivalent (up to local unitaries) to $\rho(\alpha,\theta)$, as stated above. Note also that for $k \in \mathbb{N}$ $\rho(\alpha,\theta + k \pi/4)$ is equivalent to $\rho(\alpha,\theta)$, up to local unitaries. Therefore, we can focus only on the interval $\theta \in [0,\pi/4]$.

After Alice has applied her filter, we are left with the state \eqref{Mafalda}, on which Bob will now apply his filter $F_B$. A possible approach to deal with Bob's filter is to use the concept of steering, introduced above. Indeed if a state $\rho$ is unsteerable from Alice to Bob, then the state remains unsteerable (hence local) after any local operation on Bob's side. For a proof see \cite{quintino15}, Lemma 2. In our case, this implies that if $\rho(\alpha,\theta)$ is unsteerable (from Alice to Bob), then the state
\begin{align}
 \rho_{F_B}(\alpha,\theta) =  \frac{\mathds{1} \otimes F_B \; \rho(\alpha,\theta) \; \mathds{1} \otimes F_B^{\dagger}}{\Tr(\mathds{1} \otimes F_B \; \rho(\alpha,\theta) \; \mathds{1} \otimes F_B^{\dagger})}
\end{align}
is unsteerable, hence local. Thus we can prove the theorem by showing that the state $\rho(\alpha_c,\theta)$ is unsteerable (from Alice to Bob) for all $\theta \in [0,\pi/4]$ and for all POVMs. Note that if we restricted ourselves to projective measurements on Alice's side we could use the family of LHS models presented in Ref. \cite{bowles15b}, but the requirement of general POVMs forces us to find another approach. In particular the restriction of projective measurements would prevent Alice's filter from increasing the dimension of the Hilbert space, and is thus not general enough.

In order to construct a LHS model for states of the form $\rho(\alpha,\theta)$ we use several methods, in particular the algorithmic method presented in \cite{hirsch15,cavalcanti15}. In principle this method allows us to find a LHS model for any given unsteerable state. However, here we need to prove that the entire class of states $\rho(\alpha_c,\theta)$ admits a LHS model, for a fixed value $\alpha_c$ and the whole interval $\theta \in [0,\pi/4]$. To do so we first choose finitely many angles $\theta_k \in [0,\pi/4]$ in order to get pairs $(\alpha_k,\theta_k)$ such that $\rho(\alpha_k,\theta_k)$ admits a LHS model. Then we consider convex combinations of these states with separable states in order to extend the model to the whole interval. To finish the proof, we must treat the interval boundaries, i.e. the two limits $\theta \rightarrow 0$ and $\theta \rightarrow \pi/4$, for which we use different techniques.

\begin{figure}[b!] \begin{center}
\includegraphics[width=0.9\columnwidth]{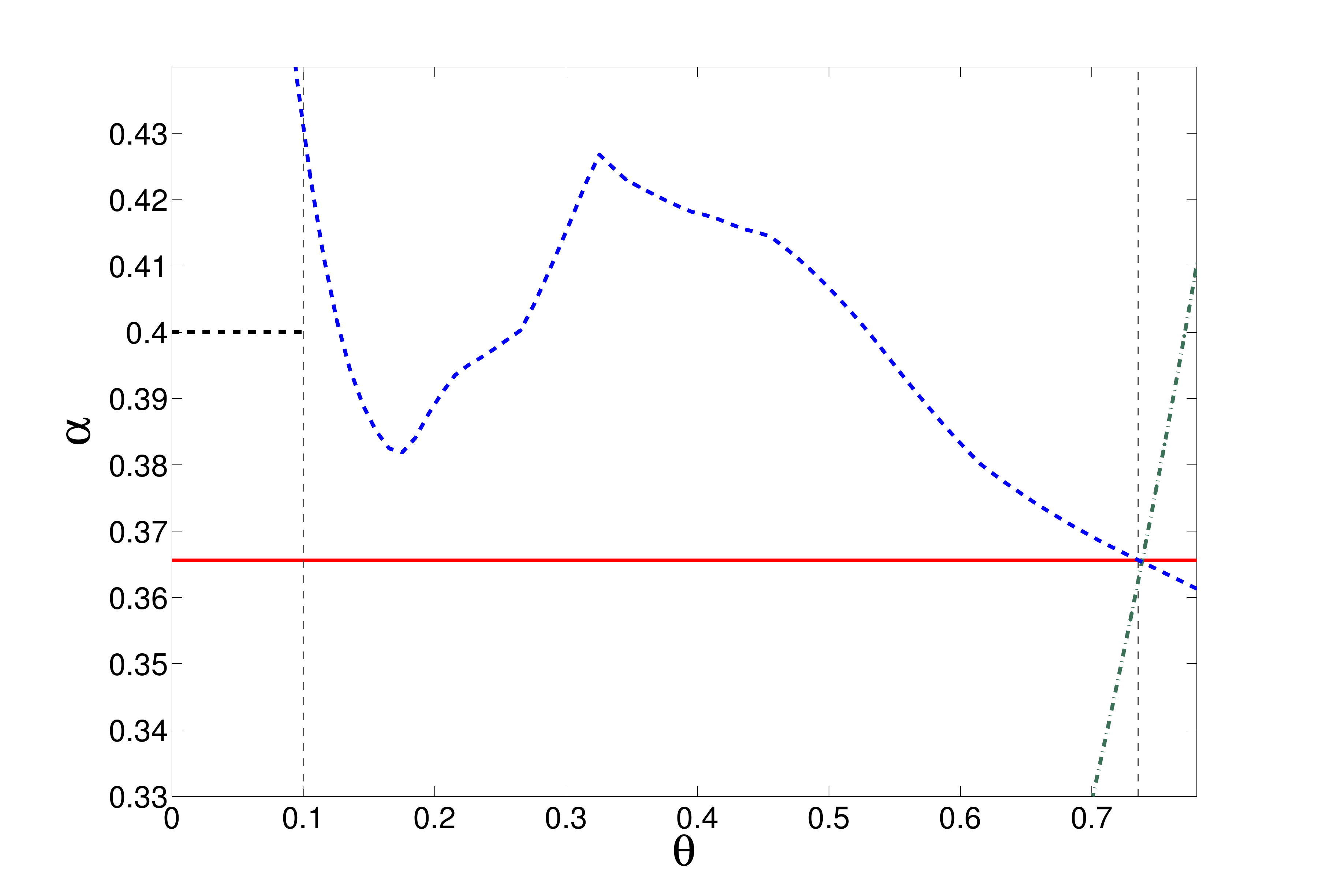}
\caption{Parameter region for which we could prove that the states $\rho(\alpha,\theta)$ of Eq. \eqref{Mafalda} admit a LHS model for all POVMs. As explained in the main text, we separate the interval $\theta \in [0,\pi/4]$ in three regions. In interval $I_1=[0,0.1]$, the existence of a LHS model is guaranteed below the black dashed horizontal line. In interval $I_2=[0.1,0.7365]$, a LHS model is demonstrated for all points below the blue dashed curve. In interval $I_3=[0.7365, \pi/4]$, a LHS model exists below the green dash-dotted curve. Overall, this guarantees that the state $\rho(\alpha,\theta)$ admits a LHS model for $\alpha \lesssim 0.3636$ and for all values of $\theta \in [0,\pi/4]$, i.e. all points below the red solid horizontal line. }
\end{center}
\end{figure}

We consequently break the interval $[0,\pi/4]$ in three sub-intervals: $I_1=[0,\theta_s]$, $I_2=[\theta_s,\theta_l]$ and $I_3=[\theta_l,\pi/4]$, where $\theta_s=0.1$ and $\theta_l=0.7365$. Let us start with $I_3$. Here $\theta$ is in the neighbourhood of $\pi/4$. We decompose the target state $\rho(\alpha,\theta)$ as a mixture of states admitting a LHS model for POVMs. Specifically, we search for which values of $\theta$ and $\alpha$, we can find a convex combination of the form:
\begin{align} \label{cd}
\rho(\alpha,\theta) = q \, \rho_W(5/12) + (1-q) \sigma
\end{align}
with $0 \leq q \leq 1$. Recall that the Werner state $\rho_W(5/12)$ admits a LHS model for POVMs \cite{barrett02,quintino15}. Here $\sigma$ is an unspecified two-qubit state, and as long as $\sigma$ admits a LHS model, this implies that $\rho(\alpha,\theta)$ is unsteerable, as one can write it as a probabilistic mixture of two unsteerable states. A simple solution is to demand that 
\begin{align} 
\sigma =  \frac{\rho(\alpha,\theta)  - q \rho_W(5/12)}{1-q} 
\end{align}
be separable. By setting $q = \frac{12}{5} \alpha  \sin(2 \theta)$, we obtain a diagonal matrix $\sigma$ (for all $\alpha$ and $\theta$). To verify that $\sigma$ represents a valid state, we only need to ensure that its eigenvalues are positive. One can check that this is the case when 
\begin{align} 
\alpha \leq \frac{1}{ (17/5) \cot{\theta} - 1 }	.
\end{align}

Now we focus on the interval $I_2$. In this regime we essentially use the technique presented in \cite{hirsch15}. More precisely we choose finitely many values $\theta_k \in I_2$ ($k=1..n$). For each of them, a slightly improved version of Protocol 1 of Ref. \cite{hirsch15} allows us to find a value $\alpha_k$ such that $\rho(\alpha_k,\theta_k)$ admits a LHS model for POVMs (more details in Appendix A). The obtained values of $\alpha_k$ and $\theta_k$ are shown on the blue dashed curve of Fig. 2. In order to extend the result to the continuous interval $I_2$, we use the following lemma, which is proven in Appendix B:

{\bf Lemma 1.} If the state $\rho(\alpha,\theta)$ is unsteerable from Alice to Bob (for POVMs) then the state $\rho(\alpha',\theta')$, with $\theta' \geq \theta$, is also unsteerable from Alice to Bob (for POVMs) as long as
\begin{align} \label{lemma1}
\tan(\theta') \frac{\alpha'}{(1+\alpha')} \leq \tan(\theta) \frac{\alpha}{(1+\alpha)}. 
\end{align}

Therefore, given a point $(\theta_k, \alpha_k)$ for which the state admits a LHS model, we can ensure that $\rho(\alpha,\theta)$ admits a LHS model as long as $\theta \geq \theta_k$ and $\alpha \leq \alpha_k \frac{\tan(\theta_k)}{\tan(\theta) (1+\alpha) - \alpha}$. As expected this is a decreasing function of $\theta$, but we only need to cover the interval $[\theta_k,\theta_{k+1}]$, implying that the minimal value of $\alpha$ in this interval is $\alpha_k  \frac{\tan(\theta_k)}{\tan(\theta_{k+1}) (1+\alpha) - \alpha}$, which is close to $\alpha_k$ if $\theta_k$ and $\theta_{k+1}$ are close. Therefore, we have to choose $\theta_{k+1}$ sufficiently close to $\theta_k$, such that the value of $\alpha$ does not drop below $\alpha_c$.

We are thus left with the interval $I_1$, where $\theta$ is in the neighbourhood of $0$. We cannot use the same method as in $I_2$ as whatever smallest $\theta_s=\min_k \{ \theta_k \}$ we choose, Lemma 1 only allows us to say something about some $\theta \geq \theta_s$, leaving the interval $[0,\theta_s]$ unsolved. Note also that by setting $\theta_s=0$ we get a separable state and consequently mixing it with another separable state cannot give rise to an entangled one (note that setting $\theta=0$ in Lemma 1, one obtains $\alpha' = 0$). 

However, in this region, an explicit LHS model for projective measurements is known \cite{bowles15b}. The model holds for $\rho(\alpha,\theta)$ as long as 
\begin{align} \label{condition} \cos(2\theta)^2 \geq \frac{2\alpha-1}{(2-\alpha) \alpha^3}.
\end{align} 
To take POVMs into account we use a method developed in Ref. \cite{hirsch13}, Protocol 2. Starting from an entangled state $\rho$ admitting a local model for projective measurements, we can construct another entangled state $\rho'$ admitting a local model for POVMs. Note that while the method was originally developed for LHV, we use it here for LHS models (i.e. only on Alice's side). Specifically, we now apply this method to the state $\rho(\alpha,\theta)$ where condition \eqref{condition} is fulfilled, thus ensuring that the state admits a LHS model for projective measurements. We obtain the class of states
\begin{align} 
\rho_{\theta} =  \frac{1}{2} ( \rho(\alpha,\theta)  + \ket{0}\bra{0} \otimes \rho_B )
\end{align}
\normalsize
where $\rho_B = \Tr_A(\rho(\alpha,\theta))$. This state is therefore unsteerable from Alice to Bob, for all POVMs. The last step consists in showing that $\rho(\alpha,\theta)$, where $\alpha > \alpha_c$ can be written as a convex combination of $\rho_\theta$ and a separable state, for all $\theta \in I_1$. This proof is given in Appendix C for $\alpha=0.4$.

Finally, we summarize these results in Fig. 2. This implies that the state $\rho(\alpha,\theta)$ is unsteerable for $\alpha \leq \alpha_c = 0.3636$, and for all $\theta \in [0,\pi/4]$. Therefore the Werner state $\rho_W( \alpha_c)$ displays no hidden nonlocality.

\section{Conclusion}

We proved that there exist entangled quantum states which do not display hidden nonlocality, i.e. which remain local after any local filtering on a single copy of the state. Specifically we showed this to be the case for some two-qubit Werner states. This consequently proves that local filters (or equivalently SLOCC procedures before the Bell test) are not a universal way to reveal nonlocality from entanglement. 

The natural question now is wether the use of even more general measurement strategies could help to reveal the nonlocality of the states we consider. In particular, one could look at sequential measurements \cite{gallego14}, beyond local filters, and consider the entire statistics of the measurements. Here, one chooses between several possible measurements (or filters) at each round. In order to show that a quantum state is local, one should now construct a LHV model that is genuinely sequential. That is, the distribution of the local variable should be fully independent of the choice of the sequence of measurements. Indeed, this is not the case in our model, as the distribution of the local variable depends on Alice's filter. Nevertheless, as our model is of the LHS form, the distribution of local states does not depend on the choice of local filter for Bob, and more generally covers any possible sequence of measurements on Bob's side. It would therefore be interesting to see if one can find a model that holds for arbitrary sequences of measurements on Alice's side as well. If this is not the case, then a sequence of measurements should be considered strictly more powerful than local filtering in Bell tests.

There exists however another possible extension of the Bell scenario, where Alice and Bob share many copies of the state. Here, in each run of the experiment, the two observers can now perform joint measurements on the many copies they hold. It has been shown that some local entangled states (in the standard Bell scenario) produce nonlocal statistics in this setup \cite{palazuelos12}. This phenomenon is known as `super-activation' of quantum nonlocality. While it is not known whether super activation is possible for all entangled states, it was nevertheless shown that any entangled state useful for teleportation (or equivalently, with entanglement fraction greater than $1/d$ where $d$ is the local Hilbert space dimension) can be super activated \cite{cavalcanti12}. 
In fact, it turns out that the class of states we considered, i.e. two-qubit Werner states, are always useful for teleportation \cite{horodecki98} and can thus be super activated. Our result thus demonstrates that quantum nonlocality via local filtering or many-copy Bell tests are inequivalent. 

Finally, the main open question is still whether there exists an entangled state that would display no form of nonlocality, considering arbitrary sequential measurements on many copies, or in quantum networks \cite{cavalcanti10,sen03} where stronger notions of nonlocality could be considered \cite{branciard09,rosset15,chaves15}. 

\section*{Acknowledgements}
 We acknowledge financial support from the Swiss National Science Foundation (grant PP00P2\_138917, Starting grant DIAQ) and from the Hungarian National Research Fund OTKA (K111734).

\phantom{A}

%%%%%%%%%%%%%%%%%%%%%%%%%%%%%%%%   End Main Text
%%%%%%%%%%%%%%%%%%%%%%%%%%%%%%%%   Begin Bibliography

%\bibliographystyle{linksen}  %Bibliography style for texts in english (Author: Mateus Araujo maltusan@gmail.com)
%\bibliography{mtqbib}        %MTQ personal bibliography database
%
\providecommand{\href}[2]{#2}\begingroup\raggedright\endgroup

%\pagebreak
\begin{appendix}

\section{Details about the algorithmic construction of LHS models}

%\begin{table}[t!] \label{table1}
%\begin{tabular}{| c || c|c|c|c|}
%  \hline
% k  & 1 & 2 & 3 & 4   \\
% \hline \hline
% $\alpha_{k}$ & 0.3624 & 0.3626 & 0.4018 & 0.8466 \\
% $\theta_{k}$ & 0.6823 & 0.5712 & 0.2935 & 0.0159 \\
% \hline
%\end{tabular} \caption{The $\alpha -\theta$ coordinates of the points representing states admitting LHS models }
% \end{table}

As stated in the main text we used the algorithmic construction of \cite{hirsch15} to find LHS models for states $\rho(\alpha,\theta)$. More precisely we note that for a fixed $\theta=\theta_f$ the state is linear with respect to $\alpha$ and we can thus use Protocol 1 of \cite{hirsch15} to find an $\alpha=\alpha_f$ such that $\rho(\alpha_f,\theta_f)$ admits a LHS model. 

We have run a slightly improved version of Protocol 1, which requires to choose a finite set of measurements $\{M _{a|x}\}$, a quantum state $\xi_A$, and to run the following SDP:

{\bf Protocol 1.} (improved version)
\begin{align} \text{find  } & q^* = \max  q  \\
\text{s.t.  }   & Tr_A(M _{a|x} \otimes \mathbb{I} \, \chi) = \sum_\lambda \sigma_\lambda D_\lambda(a|x)  \,\,  \forall a,x,  \,\,  \sigma_\lambda \geq 0 \,\,  \forall \lambda \nonumber  \\ \nonumber
&  \rho(q,\theta_f)- \eta \chi + (1-\eta) \xi_A \otimes \chi_B \geq 0 \\ \nonumber
&  (\rho(q,\theta_f)- \eta \chi + (1-\eta) \xi_A \otimes \chi_B)^{T_B} \geq 0 \\ \nonumber
& Tr(\chi) \geq 0
\end{align}
where $^{T_B}$ stands for the partial transposition on Bob's side and the optimization variable are (i) the positive matrices $\sigma_\lambda$ and (ii) a $d \times d$ hermitian matrix $\chi$ . This SDP must be performed considering all possible deterministic strategies for Alice $D_\lambda(a|x)$, of which there are $N= (k_A)^{m_A}$ (where $m_A$ denotes the number of measurements of Alice and $k_A$ the number of outcomes); hence $\lambda = 1,...,N$. 

For the answer to hold (i.e. ensuring that $\rho(q^*,\theta_f)$ admits a LHS model) the parameter $\eta$ must be smaller or equal to the `shrinking factor' of the set of all POVMs with respect to the finite set  $ \{M_{a|x} \}$ (and given state $\xi_A$), that is, the largest $\eta = \eta^*$  such that any shrunk POVM $\{M^{\eta}_a\}$ with elements defined by
\begin{align}
M_{a}^\eta = \eta M_{a} + (1-\eta) \Tr[\xi_A M_a] \mathds{1}
\end{align}
can be written as a convex combination of the elements of $ \{M_{a|x} \}$, i.e. $M_{a}^\eta = \sum_{x} p_x M_{a|x} $ ($\forall a$) with $\sum p_x=1$ and $p_x\geq 0$. The exact value $\eta^*$ is in general hard to evaluate, but Ref. \cite{hirsch13} gives a general procedure to obtain arbitrary good lower bounds on $\eta^*$, which is therefore enough for us to make sure that $\eta \leq \eta^*$.

The method requires the choice of a finite set of measurements $\{M _{a|x}\}$ which should `approximate well' the set of all POVMs. We considered a set of projective measurements, the directions of which were given by the vertices of the icosahedron, that is, 12 Bloch vectors forming an icosahedron. More precisely we consider all relabellings of $ \{ P_+ , P_- , 0 , 0 \}$ for $P_+$ being a projector onto a vertex of the icosahedron and $P_-$ onto the opposite direction. In addition we consider the four relabellings of the trivial measurement $ \{\mathds{1}_2  , 0 , 0 , 0 \}$, which comes for free as it cannot help to violate any steering or Bell inequalities and consequently does not even need to be inputed in Protocol 1. The set thus have 76 elements, but we need to take into account only 6 of them when running the Protocol, corresponding to the vertices in the upper half sphere of the icosahedron.

There is one degree of freedom left: the quantum state $\xi_A$. We thus computed lower bounds on the shrinking factors for different $\xi_A$ of the form $\xi_A= p \ket{0}\bra{0} + (1-p)\mathds{1}/2 $. The different estimates for $\eta^*$ in function of $p$ are given in Table 1, and the best points obtained are shown on Fig. 2.

\begin{table}[t!] \label{table1}
\begin{tabular}{| c || c|c|c|c|c|c|c|c|c|c|}
  \hline
 p  & 0 & 0.1 & 0.2 & 0.3 & 0.4 & 0.5 & 0.6 & 0.7 & 0.8 & 0.9   \\
 \hline
 $\eta$ & 0.67 & 0.67 &   0.66 &   0.66 &  0.66  &  0.66  &  0.62 &   0.56 &   0.47  &  0.32 \\
 \hline
\end{tabular} \caption{Lower bounds on shrinking factors of the set of qubit POVMs with respect to the icosahedron and with $\xi_A= p \ket{0}\bra{0} + (1-p)\mathds{1}/2 $  }
 \end{table}

\section{Proof of Lemma 1}

If the state $\rho(\alpha,\theta)$ is unsteerable from Alice to Bob (for POVMs) then the state $\rho(\alpha',\theta')$, with $\theta' \geq \theta$, is also unsteerable from Alice to Bob (for POVMs) as long as
\begin{align} \label{lemma1_appendix}
\tan(\theta') \frac{\alpha'}{(1+\alpha')} \leq \tan(\theta) \frac{\alpha}{(1+\alpha)}. 
\end{align}

To prove this lemma we show that $\rho(\alpha',\theta')$ can be written as a convex combination of $\rho(\alpha,\theta)$ and a separable state, as long as condition \eqref{lemma1_appendix} holds. We want:

\begin{align}
\rho(\alpha',\theta') = q \rho(\alpha,\theta) + (1-q) S
\end{align}
where $S$ is a separable state. Inverting this relation we get: 
\begin{align}
(1-q) S = (\rho(\alpha',\theta')- q \rho(\alpha,\theta) ).  
\end{align}
That is:

\tiny
\begin{align}
(1-q) S = \frac{1}{2}\begin{pmatrix} \nonumber
\text{cos}^2\theta' (1+\alpha') - q \text{cos}^2\theta (1+\alpha) & 0 & 0 & \alpha' \text{cos}\theta' \text{sin}\theta' - q \alpha \text{cos}\theta \text{sin}\theta  \\
0 & \text{cos}^2\theta' (1-\alpha') - q \text{cos}^2\theta (1-\alpha) & 0 & 0 \\
0 & 0 & \text{sin}^2\theta' (1-\alpha') - q \text{sin}^2\theta (1-\alpha) & 0 \\
\alpha' \text{cos}\theta' \text{sin}\theta' - q \alpha \text{cos}\theta \text{sin}\theta & 0 & 0 &\text{sin}^2\theta' (1+\alpha') - q \text{sin}^2\theta (1+\alpha) 
\end{pmatrix} \nonumber
\end{align}
\normalsize

Setting $q = \frac{\alpha' \text{cos}\theta' \text{sin}\theta'}{\alpha \text{cos}\theta \text{sin}\theta}$ makes $S$ diagonal, thus separable. We need $0 \leq q \leq 1$, that is:

\begin{align}\label{lambda}
\alpha' \leq \frac{\alpha \text{cos}\theta \text{sin}\theta}{ \text{cos}\theta' \text{sin}\theta'}	.
\end{align}
Under this condition we are then left to show that $S$ is a valid state, i.e. a semi-definite positive trace-one matrix. First one can check that its trace is always equal to $1$, second its eigenvalues are just given by the diagonal elements, this gives us the four following conditions for positivity:
\begin{align}
\alpha' (\text{cos}^2\theta' - \ell \text{cos}^2\theta (1+\alpha) ) + \text{cos}^2\theta' \geq 0 \label{ineq1} \\ 
\alpha' (-\text{cos}^2\theta' - \ell \text{cos}^2\theta (1-\alpha) ) + \text{cos}^2\theta' \geq 0 \label{ineq2}  \\
\alpha' (-\text{sin}^2\theta' - \ell \text{sin}^2\theta (1-\alpha) ) + \text{sin}^2\theta' \geq 0 \label{ineq3}  \\
\alpha' (\text{sin}^2\theta' - \ell \text{sin}^2\theta (1+\alpha) ) + \text{sin}^2\theta'\geq 0 \label{ineq4} 
\end{align}
where $\ell = \frac{ \text{cos}\theta' \text{sin}\theta'}{\alpha \text{cos}\theta \text{sin}\theta}$. Each inequality is of the form $A \alpha' + C \geq 0$ with $C$ always positive. if $A \geq 0$ the inequality is satisfied (since $\alpha' \geq 0$) so the non-trivial case corresponds to $A < 0$ leading to the solution $ \alpha' \leq - \frac{C}{A}$. Let us focus on inequalities \eqref{ineq2} and \eqref{ineq3}. first. We have:
\begin{align}
\eqref{ineq2} \implies \alpha' \leq \frac{1}{ 1 + \ell_1 (1-\alpha)} \hspace{1cm} \eqref{ineq3} \implies \alpha' \leq \frac{1}{ 1 + \ell_2 (1-\alpha)}
\end{align}
where $\ell_1 = \frac{ \text{cos}\theta \text{sin}\theta'}{\alpha \text{cos}\theta' \text{sin}\theta}$, $\ell_2 = \frac{ \text{cos}\theta' \text{sin}\theta}{\alpha \text{cos}\theta \text{sin}\theta'}$. 

Now we use the fact that $\theta' \geq \theta$ to get that $\text{sin}\theta' \geq \text{sin}\theta$ and $\text{cos}\theta \geq \text{cos}\theta'$, in the range we are interested in: $\theta \in [0,\pi/4]$. This implies $\ell_1 \geq 1/\alpha \geq 1$ while $\ell_2 \leq 1/\alpha $ and $\ell_1 \geq \ell_2$ meaning that the inequality \eqref{ineq2} is always more constraining than the inequality \eqref{ineq3} in the range of interest. 

We can similarly merge the inequalities \eqref{ineq1} and \eqref{ineq4} We have 
\begin{align}
\eqref{ineq1} \implies \alpha' \leq \frac{1}{\ell_1 (1+\alpha) - 1} \hspace{1cm} \eqref{ineq4} \implies \alpha' \leq \frac{1}{\ell_2 (1+\alpha) - 1}
\end{align}
and again using that $\ell_1 \geq \ell_2$ we see that that the inequality \eqref{ineq1} is more constaining than the inequality \eqref{ineq4}. We are thus left with the two inequalities \eqref{ineq1} and \eqref{ineq2}, but one can show that the inequality \eqref{ineq1} is more constraining. One has 
\begin{align}
\frac{1}{\ell_1 (1+\alpha) - 1} &\leq \frac{1}{ 1 + \ell_1 (1-\alpha)} \\ 
\ell_1 (1+\alpha) - 1 &\geq 1 + \ell_1 (1-\alpha) \nonumber \\ 
2\alpha\ell_1 &\geq 2  \nonumber
\end{align}
which is true since $\ell_1 \geq 1/ \alpha$. Finally we can show that the inequality \eqref{ineq1} is more constraining than condition \eqref{lambda}:
\begin{align}
\frac{1}{\ell_1 (1+\alpha) - 1} &\leq \frac{1}{\ell} \\ \nonumber
\ell_1 (1+\alpha) &\geq 1 + \ell \\ \nonumber
1 + \alpha &\geq \ell/\ell_1 + 1/\ell_1 .
\end{align}
To see that the last inequality is true one can compare terms by terms: we have that  $\ell/\ell_1 =  \frac{\text{cos}^2\theta'}{\text{cos}^2\theta} \leq 1$ and $1/\ell_1 \leq \alpha$, once again using $\theta' \geq \theta$ and $\theta \in [0,\pi/4]$.

\section{POVM model for small $\theta$ }

Here we give the proof that for $\theta \in [0,0.1]$ the state $\rho(0.4,\theta)$ can be written as a convex combination of a separable state and $\rho_{\theta} =  \frac{1}{2} ( \rho(\beta,\theta)  + \ket{0}\bra{0} \otimes \rho_B )$, where $\rho_B = \Tr_A(\rho(\beta,\theta))$ and $\beta$ and $\theta$ are linked by $\cos(2\theta)^2 \geq \frac{2\beta-1}{(2-\beta) \beta^3}$.

We want:
\begin{align}
\rho(\alpha,\theta) = q \rho_{\theta} + (1-q) S
\end{align}
where $S$ is a separable state. Inverting this relation we get: 
\begin{align}
(1-q) S = (\rho(\alpha,\theta)- q \rho_{\theta} ) 
\end{align}
where the non-zero elements of $S$ are given by:
\begin{align} 
&S(1,1)=2\text{cos}^2\theta (1+\alpha) - q (2\text{cos}^2\theta (1+\beta) + \text{sin}^2\theta (1-\beta)) \\ 
&S(2,2)=2\text{cos}^2\theta (1-\alpha) - q (2\text{cos}^2\theta (1-\beta) + \text{sin}^2\theta (1+\beta)) \\
&S(3,3)=2 \text{sin}^2\theta (1-\alpha) - q \text{sin}^2\theta (1-\beta) \\ 
&S(4,4)=2 \text{sin}^2\theta (1+\alpha) - q \text{sin}^2\theta (1+\beta)  \\
&S(1,4)=S(4,1)=4 \alpha \text{cos}\theta \text{sin}\theta - q 2 \beta \text{cos}\theta \text{sin}\theta 	.
\end{align}

To prove that $S$ can be made separable we set $\alpha=0.4$, $q=1/2$ and $\beta=3/4$, which leads to $\theta \lesssim 0.11$, for which values the matrix $S$ can be proved positive and separable (via PPT criterion \cite{horodecki_ppt}). To extend it one just has to note that the positivity and the PPT constraints of $S$ are of the form $\text{sin}^2\theta (A\text{cos}^2\theta - B\text{sin}^2\theta) \geq 0$, where $A,B \geq 0$. This implies that if $S$ is positive and separable for $\theta$, so is it for $\theta' \leq \theta$, for any $\theta \in [0,0.1]$. 

\end{appendix}

\end{document}